\documentclass[letterpaper,twocolumn,10pt]{article}
\usepackage{./usenix2019_v3}

\usepackage{tikz}
\usepackage{amsmath}
\usepackage{setspace}
\usepackage{xcolor}

\usepackage{filecontents}
\usepackage{comment}
\usepackage{multirow}
\usepackage{colortbl}
\usepackage{pifont}

\usepackage{enumitem}

\usepackage{kotex}
\newcommand{\sgxssd}{\texttt{SGX-SSD}}
\newcommand{\squishlist}{
	\begin{list}{$\bullet$}
		{ \setlength{\itemsep}{0pt}      \setlength{\parsep}{-0pt}
			\setlength{\topsep}{4pt}       \setlength{\partopsep}{0pt}
			\setlength{\listparindent}{-2pt}
			\setlength{\itemindent}{-5pt}
			\setlength{\leftmargin}{1em} \setlength{\labelwidth}{0em}
			\setlength{\labelsep}{0.5em} } }
	
\newcommand{\squishend}{
\end{list}  }

\newcommand{\myintertext}[1]{\end{enumerate} #1}
\newcommand{\myintertextend}{\begin{enumerate}[resume]}
\newcommand{\boldtitle}[1]{\textbf{#1}\hspace{0.3cm}}

\begin{document}

\date{}

\title{{\sgxssd}: A Policy-based Versioning SSD with Intel SGX}


\author{
     {\rm Jinwoo Ahn$^{1}$, Seungjin Lee$^{1}$, Jinhoon Lee$^{1}$, Yungwoo Ko$^{1}$, Donghyun Min$^{1}$}\\
     {\rm Junghee Lee$^{2}$, Youngjae Kim$^{1}$}\\
$^{1}$Sogang University, $^{2}$Korea University, Seoul, Republic of Korea\\
}
\maketitle

\setstretch{0.964}
\sloppy

\begin{abstract}
This paper demonstrates that SSDs, which perform device-level versioning, can be exposed to data tampering attacks when the retention time of data is less than the malware's dwell time.
To deal with that threat, we proposed {\sgxssd}, a SGX-based versioning SSD which selectively preserves file history based on the given policy.
The proposed system adopts Intel SGX to implement the version policy management system that is safe from high-privileged malware. 
Based on the policy, only the necessary data is selectively preserved in SSD that prevents files with less priority from wasting space and also ensures the integrity of important files.
\end{abstract}

\vspace{-0.15in}
\section{Introduction}
\label{sec:intro}
\vspace{-0.1in}


Recently, malware corruption attacks are becoming more astute, such as ransomware~\cite{ransomware1, ransomware2, ransomware_many, intelligent_ransomware} and wiper~\cite{wiper, wiper_petya, wiper_shammon_1, wiper_shamoon_2}. 
Most of them use the latest exploit techniques to escalate privilege~\cite{vulnerableFS} and compromise the system~\cite{ransomware_many, wannacry, sodinokibi}. 
Moreover, to eliminate the possibility of data restoration, they neutralize the existing software-based backup system by performing attacks aiming the software backup device that is connected locally or remotely~\cite{ransomware_many,ryuk1,ryuk2}.


To overcome this problem, recent studies have moved their attention from the host side software-based backup system, 
which can be compromised,
to safely versioning the data inside a device~\cite{amoeba,SSDinsider,flashguard,almanac,ransomblocker}. 
These studies commonly protect the integrity of data by retaining the past version of data inside SSD\footnote{In this paper, this device is called a versioning SSD.}.
Since SSD firmware is isolated from the host, compromising the versioning system inside SSD is almost impossible even though malware escalates its privilege~\cite{flashguard, almanac}.


Project Almanac~\cite{almanac} is a state-of-art versioning SSD that preserves all the data updated by a user for a certain period. 
However, Project Almanac has a critical security issue. 
To explain this problem in a more comprehensive way, we define the terms pertaining to versioning in Table~\ref{tab:Declare}. 
In Project Almanac, every data page shares the same Retention Time(\texttt{RT}). 
Thus, when the RT value is altered by Project Almanac, all pages will get the newly defined \texttt{RT} value. 
Since every page has the same \texttt{RT} value, important files and less important files will have the same \texttt{RT}, which is a limitation of the system. 
Too high \texttt{RT} value leads to higher space overhead due to versioning, while too low \texttt{RT} value introduces a risk of data integrity violation by malware. 
Project Almanac can dynamically control the \texttt{RT} by the predefined algorithm, while guaranteeing at least 3 days of \texttt{RT} to prevent data retention time from becoming too short.
When SSD lacks of space due to too many old versions (\texttt{OV}), Project Almanac implements the retrieving mechanism that reclaims the older ones.

\begin{figure}[!t]
		
	\begin{center}
		\begin{tabular}{@{}c@{}c@{}c@{}c@{}}
			\includegraphics[width=0.3\textwidth]{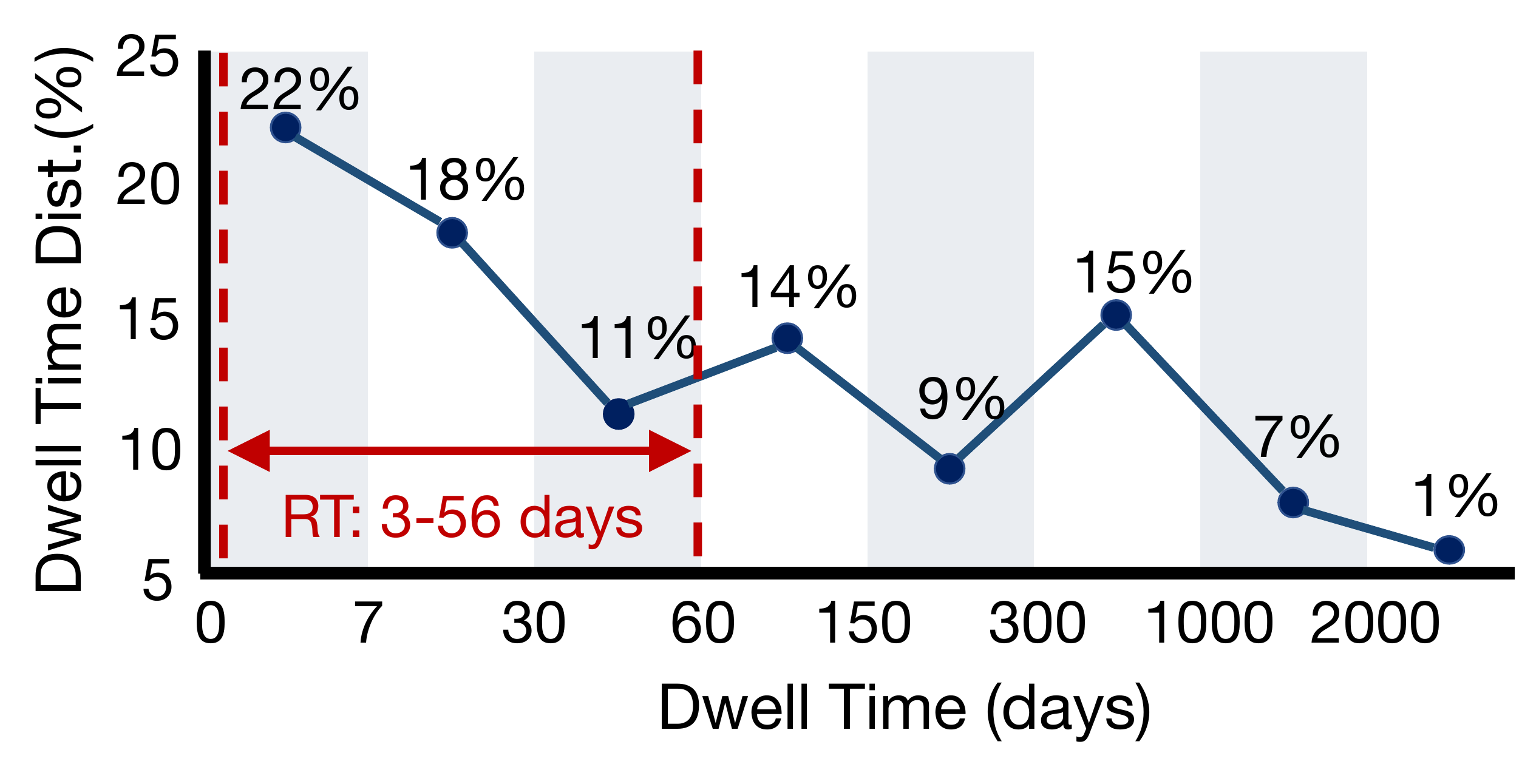} 
		\end{tabular}
		\vspace{-0.2in}
		\caption{\small Global Dwell Time Distribution of Malware~\cite{dwelltime}.
			}	\label{fig:dwelltime}
		\vspace{-0.4in}
	\end{center}
\end{figure}


Figure~\ref{fig:dwelltime} shows the statistical distribution of malware Dwell Time (\texttt{DT}). 
As shown in the graph, over 50\% of attacks have 60 days or longer \texttt{DT}. 
In Project Almanac, however, data integrity is not guaranteed if \texttt{DT} is greater than \texttt{RT}.
For instance, malware can execute a Delayed Attack\footnote{The Delayed Attack~\cite{inuksuk} is a data corruption attack that aims the data's \texttt{OV} to expire, which cannot be restored.}.
Assume that ransomware successfully evades detection and breaks into a host system, and encrypts the data that has 3 days of \texttt{RT}.
Afterward, malware can circumvent detection and stay for 4 days (\texttt{DT}=4days). 
During that time, when the user requests to read the data, it deceives the user by decrypting the encrypted data and handing it over to the user. 
At the 4th day, the user will realize the lost data, which is too late for the user, and request data restoration to Project Almanac. However, since the original data in Project Almanac is reclaimed due to expired \texttt{RT}, there is no way for the user other than paying a ransom.


{\bf Main Idea:} 
The essence of Project Almanac's problem is that the space consumption of the \texttt{OV} increases to provide the same \texttt{RT} for all data.
To mitigate this, the minimum \texttt{RT} is set short, but the integrity of all data 
cannot be guaranteed 
for malware with 3 days or longer \texttt{DT}. 
To solve this problem, we propose {\sgxssd}, which is a policy-based file versioning SSD. 
In SSD firmware, rather than applying the same policy to all files, the version policy is selectively applied to the files based on the importance specified by the user.
{\sgxssd} consists of following two important software/hardware components(\texttt{SPM}, \texttt{PV-SSD}).

\squishlist
\item
SGX-based Secure Policy Manager(\texttt{SPM}): 
It is a SGX-based policy manager for the safe policy management. Trusted Communication Channel is implemented to safely transfer user configurations to SSD.
\item
Policy-based File Versioning SSD(\texttt{PV-SSD}): 
SSD does not have a file semantic inside. 
The policy-based file versioning SSD perceives the file semantic to perform file-based versioning and restoration based on the policy.
\squishend

\begin{table}[!t]
	\centering	
	\small
	\begin{center}
		\resizebox{\columnwidth}{!}{
			\begin{tabular}{|l||p{7cm}|c|c|c|c|c|}
				\hline
				\textbf{Term}&\textbf{Description}\\
				\hline
				\hline
				Retention Time ({\texttt{RT}})&Period to guarantee recovery after data becomes invalid\\
			    \hline
			    \multirow{2}{*}{Dwell Time ({\texttt{DT}})}&The period of time between malware executing within an environment and it being detected\\
			    \hline
			    Old Version ({\texttt{OV}})&Previous version of data that is preserved for recovery\\
			    \hline
		\end{tabular}}
	\vspace{-0.15in}
    	\caption{\small Definition of Terms for Versioning Systems.}
	    \label{tab:Declare}
	\vspace{-0.42in}
	\end{center}
\end{table}

{\sgxssd} removes unnecessary waste of space that results from preserving the history of low-priority files. By using the secured space, it guarantees the integrity of important files by preserving file history with safe policy.

\vspace{-0.15in}
\section{Background}
\label{sec:back}
\vspace{-0.1in}

\subsection{Intel SGX and Versioning SSD}
\vspace{-0.05in}
Intel SGX~\cite{costan2016intel} is an instruction set provided by Intel Processor of Skylake or higher. 
Through the Trusted Execution Environment (TEE) called Enclave, SGX ensures the confidentiality and integrity of applications even when the OS is compromised.
The software developer divides the application into untrusted parts and enclave parts, and implements the interface between the two parts.
The untrusted part calls the enclave through ECALL, and the enclave calls the untrusted part through OCALL.
For using hardware resources, since the enclave cannot directly call system calls, it should execute OCALL first and request the untrusted part to call the system call.
Accordingly, when the OS is compromised, there is a limitation that  SGX cannot trust user inputs through the user interface (UI).
To overcome this, Aurora~\cite{aurora} uses System Management Mode (SMM), another privileged mode provided by the processor, to ensure secure I/O between the enclave and UI.


A NAND flash-based SSD~\cite{SSD} 
logs user data to a physical address in hardware.
Since the SSD firmware is isolated from the host system, privileged malware cannot read or tamper with it~\cite{almanac, flashguard, firmware_isolated}.
In addition, it is less vulnerable from malware because it has a much smaller trusted computing base (TCB) compared to the OS kernel and other software-based solutions~\cite{almanac, flashguard}.
Versioning SSDs~\cite{flashguard, SSDinsider, amoeba, ransomblocker, almanac} utilize these  characteristics of SSDs to perform safe versioning from high-privileged malware attacks.
In particular, by utilizing the out-of-place update characteristics of the SSD, they create an old version inside the SSD without additional copying, so no additional device is required.

\vspace{-0.15in}
\subsection{Motivation: Integrity Vulnerability}
\vspace{-0.1in}
When using a versioning SSD such as Project Almanac, the user requests data recovery from the SSD only after malware is detected.
In this environment, if malware having \texttt{DT} longer than the minimum \texttt{RT} of Project Almanac (default: 3 days) invades the host system, Project Almanac can not guarantee the integrity of the whole storage data.
The reason is that at the time of detecting the malware, the \texttt{RT} of the \texttt{OV} has already expired, so the \texttt{OV} can not be recovered. 

\begin{figure}[!t]
	\centering
    
    \begin{tabular}{c}
        \includegraphics[width=0.32\textwidth]{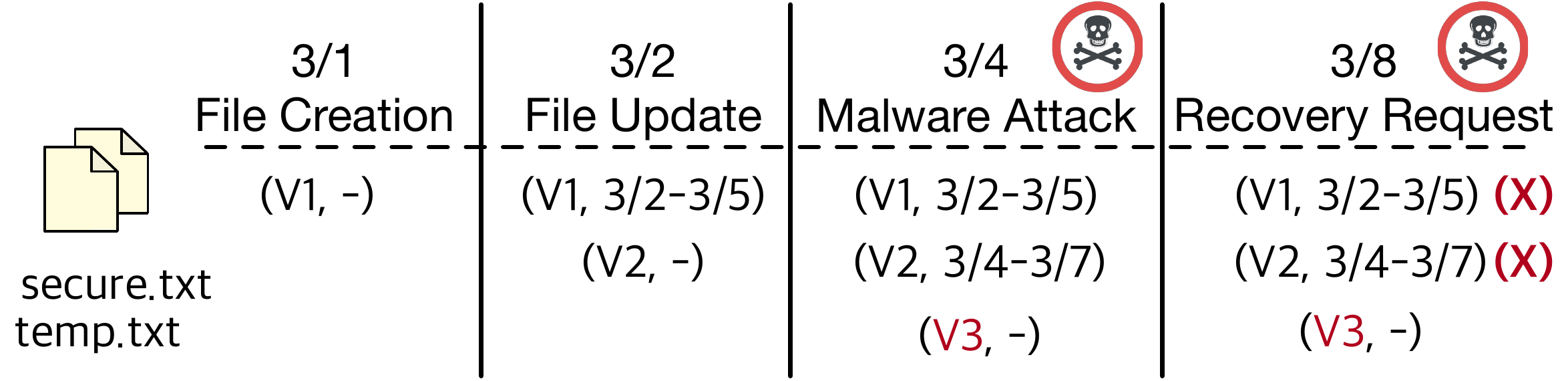}\\
        \small (a) Delayed Attack in Project Almanac\\
            \includegraphics[width=0.32\textwidth]{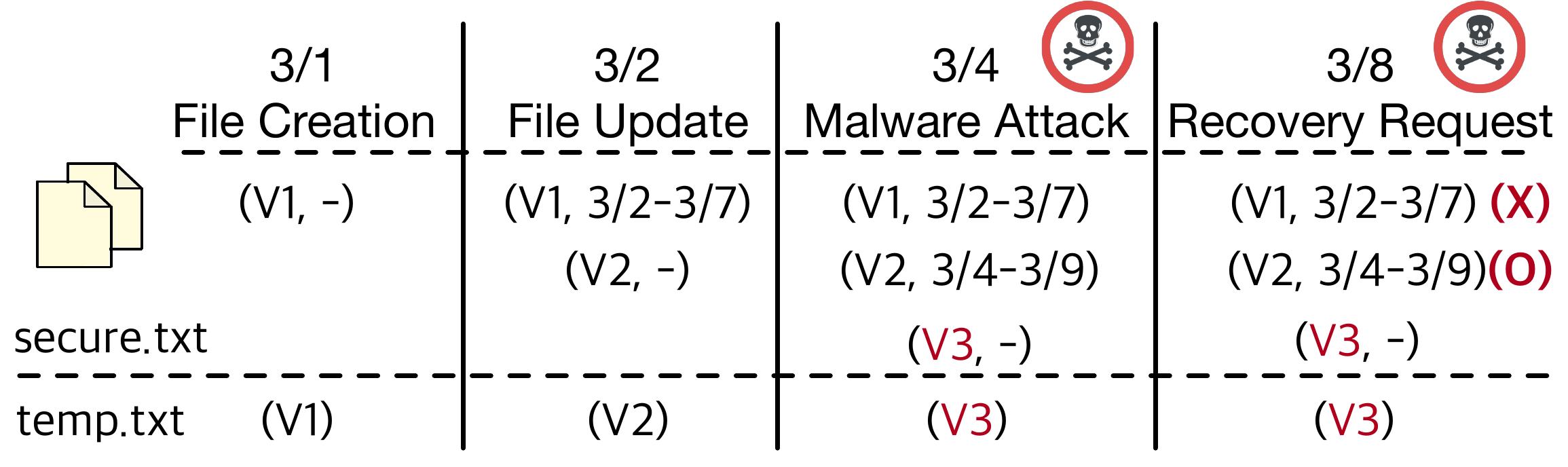}\\
        \small (b) Delayed Attack in {\sgxssd}\\
        \end{tabular}
        \vspace{-0.1in}
      \caption{\small 
      Examples of Delayed Attack Scenarios.
      }
	\label{fig:delayed_attack}
	\vspace{-0.2in}
\end{figure}

Figure~\ref{fig:delayed_attack}(a) shows a scenario in which malware performs a delayed attack in Project Almanac.
Suppose {\texttt{secure.txt}} is an important file to be protected by the user, and {\texttt{temp.txt}} is a temporary file that does not need to be protected.
Assume that the malware's {\texttt{DT}} is 4 days, and that Project Almanac's minimum {\texttt{RT}} is 3 days.
In addition, it is assumed that the host system has a high I/O intensity, so that the Project Almanac's \texttt{RT} converges to the minimum \texttt{RT}.
Also, malware can intentionally perform a lot of I/O to lower Project Almanac's \texttt{RT} value.
The user creates files ({\texttt{secure.txt}} and {\texttt{temp.txt}}) on 3/1 in the Project Almanac environment and updates the files on 3/2.
At this point, the file data (V1) generated on 3/1 becomes the {\texttt{OV}}, and the newly updated data (V2) becomes valid data.
In 3/4, malware with a {\texttt{DT}} of 4 days invades the host system and encrypts the files ({\texttt{secure.txt}} and {\texttt{temp.txt}}).
At this time, there are two {\texttt{OV}} (V1, V2) on the SSD, and the data (V3) encrypted by malware becomes valid data.
{\texttt{RT}} of V1 ends at 3/5, and {\texttt{RT}} of V2 ends at 3/7, and V1 and V2 are erased.
The user detects malware at 3/8 and tries to recover data.
However, since the data (V1, V2) stored before the malware invasion does not remain on the SSD, both \texttt{secure.txt} and \texttt{temp.txt} fail to be recovered.

Figure ~\ref{fig:delayed_attack}(b) describes a scenario where malware performs a delayed attack in {\sgxssd}.
It is assumed that the user sets the {\texttt{RT}} of the important {\texttt{secure.txt}} to 5 days,
and {\texttt{RT}} is not set in {\texttt{temp.txt}} because its importance is low.
The user updates the data with the flow described above, and the malware attempts to encrypt it in the same way.
At 3/8 when malware is detected, the user requests data recovery.
{\texttt{temp.txt}} cannot be recovered because there is no \texttt{OV}.
On the other hand, \texttt{secure.txt} has \texttt{RT} of 5 days, and V2 has not yet expired.
Therefore, the user can recover \texttt{secure.txt} to the point in time before the malware entered.
In addition, on 3/8, Project Almanac has two versions of \texttt{secure.txt} and \texttt{temp.txt}, so there are a total of four \texttt{OV}.
On the other hand, {\sgxssd} does not have the \texttt{OV} of \texttt{temp.txt}, so it only has 2 \texttt{OV}s. 
This shows that the space resources obtained by reducing the \texttt{RT} of the less important file (\texttt{temp.txt}) can be used to protect the important file (\texttt{secure.txt}) for a longer period.

The characteristic of Project Almanac is that it is an SSD that preserves the state (State-Preserving SSD).
The limitation of this is that all blocks have the same RT. 
We aim to support secure, selective file versioning that is not vulnerable to such malware attacks.

\vspace{-0.15in}
\section{Threat Model}
\label{sec:threat}
\vspace{-0.15in}


Malware infiltrates the host system with software to obtain the user, root, or highest privilege (ring-0 level). 
Malware aims to delete or tamper with sensitive data of user. We exclude attacks through physical access from our threat model.
Against the threat of malware, {\sgxssd} guarantees the integrity of the all old versions of data stored on disk based on the policy.
However, {\sgxssd} does not guarantee the integrity of newly updated data after malware infection.
What we guarantee is limited to pre-existed data that was updated to the disk before malware infection.
Through this, when the data is tampered with, the user recovers the tampered files to the point in time before the malware invades.
Also, according to the policy, the integrity of files that have already expired is not guaranteed.
Since the user recognizes this, one can allocate a policy with sufficiently longer \texttt{RT} to protect data if the file is important.
Finally, since SGX and SSD each has a TEE and a very small TCB, it is assumed to be safe from malware.
Also, the user trusts the vendor that distributes {\sgxssd}, and assumes that the user's system is not infected with malicious code when the user first installs {\sgxssd}.
\vspace{-0.15in}
\section{Design and Implementation}
\label{sec:design}
\vspace{-0.15in}


To design a secure policy-based versioning SSD even in a compromised OS environment, we answer two questions:

\squishlist
\item
The user's request inevitably passes through an untrusted OS where the request can be tampered with by a man-in-the-middle attack.
\textit{How can we guarantee that the policy request entered by the user is safely delivered to the SSD?}
\item
SSD is a block device, so there is no file semantics, whereas policies are assigned on a file-by-file basis.
\textit{In order for the SSD to perform policy-based versioning at the block level, how can the SSD recognize the file semantics corresponding to each block?}
\squishend


{\sgxssd} is composed of software module called SGX-based Security Policy Manager (\texttt{SPM}) and a hardware module called Policy-based File Versioning SSD (\texttt{PV-SSD}).
When the vendor deploys {\sgxssd}, the vendor hardcodes and hides the unique device key ($K_{dev}$) in each module (\texttt{SPM} and \texttt{PV-SSD} firmware).
$K_{dev}$ is a secret shared between two modules.



\begin{table}[!b]
	\vspace{-0.2in}
	\centering	
	\small
	\begin{center}
		\resizebox{\columnwidth}{!}{
			\begin{tabular}{|l||p{7cm}|c|c|c|c|c|}
				\hline
				\textbf{Notation}&\textbf{Description}\\
				\hline
				\hline
				$K_{dev}$&Private key shared by \texttt{SPM} and \texttt{PV-SSD}.\\
				\hline
			    $f_j$&File to apply the policy.\\
			    \hline
			    $P_{(f_{j})}$&The policy setting value entered by user. It consists of the command type (\texttt{CREATE}, \texttt{CHANGE}, \texttt{DELETE}), the file path of $f_j$, and \texttt{CP} (\texttt{CP} is NULL when the command type is \texttt{DELETE}.).\\
			     \hline
                $pbSet_{(f_{j})}$&Piggyback set sent by the file system. It consists of  the file path of $f_j$ and the offset corresponding to data block.\\
                \hline
                $R_{(P_{(f_{j})})}$&A response message that \texttt{PV-SSD} sends to \texttt{SPM}. If the request is successful, a success message (\texttt{SUCCESS}) is sent, and if it fails, an error code is sent.\\
                \hline
                $rec_{(f_{j})}$&Message that the recovery tool sends to \texttt{PV-SSD} to request recovery. The file path and the list of all LBAs belonging to $f_j$, a certain time (\texttt{T}) or the version (\texttt{V}) to rollback are included. If the LBA list of $rec_{(f_{j})}$ is NULL, \texttt{PV-SSD} exhaustively searches all physical pages to restore the file.\\
			    \hline
			    $MAC_{K}(M)$&Generate Message Authentication Code (MAC) generated from message (M) using key (K)\\ 
			    \hline
                $E_{K}(M)$&Encrypt the message (M) using Key (K)\\
                \hline
                $D_{K}(M)$&Decrypt the message (M) using Key (K)\\
                \hline
		\end{tabular}}
	\vspace{-0.15in}
    	\caption{\small Notation for {\sgxssd}.}
	    \label{tab:notation}
	\vspace{-0.0in}
	\end{center}
\end{table}

\vspace{-0.15in}
\subsection{Policy Management via User Interface}
\vspace{-0.05in}

In {\sgxssd}, the versioning policy is a strategy on how to perform versioning on file data. 
This versioning policy consists of configuration parameters (\texttt{CP}), which is policy information, and a path of file to be managed based on the policy.
Especially, \texttt{CP} consists of Retention Time (\texttt{RT}), Backup Cycle (\texttt{BC}), the maximum number of versions (\texttt{V}). 
In addition, definitions of the terms used in the {\sgxssd} and the mathematical notations are shown in Table~\ref{tab:notation}.


The user sets a new versioning policy in {\sgxssd} in the following steps. 
First, the user inputs the following command in the console window (User Interface) to run the \texttt{SPM}; {\texttt{\$./sgx\_ssd}}.
Second, once \texttt{SPM} runs, the user creates, changes or deletes the policy by typing the following command; {\texttt{\$\{Policy Create|Change|Delete\}\{File path\}\{retention time\}\{Backup cycle\}\{Number of Version\}}}. 
If the user newly wants to create a policy, the user inputs the file path and \texttt{CP} to register with the policy.
Versioning of the file registered in the policy begins from the time the user creates the policy.
If the user wants to change the policy, the user also inputs the file path and \texttt{CP} values.
The versioning policy of the file changes from the time user changes the policy.
If the user wants to delete the policy, the user only inputs the file path. 
In this case, not only file itself but also the old versions of the file registered in the policy are deleted.
At this time, the \texttt{OV} of the file registered in the policy is deleted along with the policy.
The user can recover files to a certain time or a certain version by using {\sgxssd} recovery tool. The user inputs the following command to console; {\texttt{\$./recovery\_tool \{file path\}\{Time|Version\}}}.

\vspace{-0.15in}
\vspace{-0.05in}
\subsection{SGX-based Policy Manager}
\label{sec:spm}
\vspace{-0.05in}



Figure~\ref{fig:overview} shows the overall design and operational flow of {\sgxssd}. 
The user sets the file policy through the control path.
\texttt{SPM} safely delivers the user’s versioning policy from UI to the enclave by using the approach provided by Aurora~\cite{aurora}. 
First, when the user runs \texttt{SPM}, the \texttt{SPM} suspends the OS and switches to the SMM. 
The SMM process receives the versioning policy ($P_{(f_{j})}$) of the file ($f_j$) from the user through the keyboard.
Since the OS is suspended, even privileged malware cannot interfere with this procedure.
After the user input is finished, the user input ($P_{(f_{j})}$) is safely transmitted to the enclave through a secure session implemented in advance between the SMM process and the \texttt{SPM} enclave.
\texttt{SPM} enclave encrypts $P_{(f_{j})}$ by using $K_{dev}$ and generates the MAC ($MAC_{K_{dev}}$($E_{K_{dev}}$($P_{(f_{j})}$))).
The generated message ($E_{K_{dev}}$($P_{(f_{j})}$), $MAC_{K_{dev}}$($E_{K_{dev}}$($P_{(f_{j})}$))) is sent to \texttt{PV-SSD}.
The Policy Management (\texttt{PM}) module operating inside \texttt{PV-SSD} decrypts the message and verifies its integrity using $K_{dev}$ to retrieve $P_{(f_{j})}$.
Once the policy update ends, \texttt{PM} encrypts the response ($R_{(P_{(f_{j})})}$), generates the MAC  ($MAC_{K_{dev}}$($E_{K_{dev}}$($R_{(P_{(f_{j})})}$)), and sends the message ($E_{K_{dev}}$($R_{(P_{(f_{j})})}$), $MAC_{K_{dev}}$($E_{K_{dev}}$($R_{(P_{(f_{j})})}$))) to \texttt{SPM}. 
\texttt{SPM} uses $K_{dev}$ to decrypt and verifies the response ($R_{(P_{(f_{j})})}$).


\vspace{-0.15in}
\vspace{-0.05in}
\subsection{File System Modification}
\label{sec:os}
\vspace{-0.05in}

\begin{figure}[!t]
		
	\begin{center}
		\begin{tabular}{@{}c@{}c@{}c@{}c@{}}
			\includegraphics[width=0.42\textwidth]{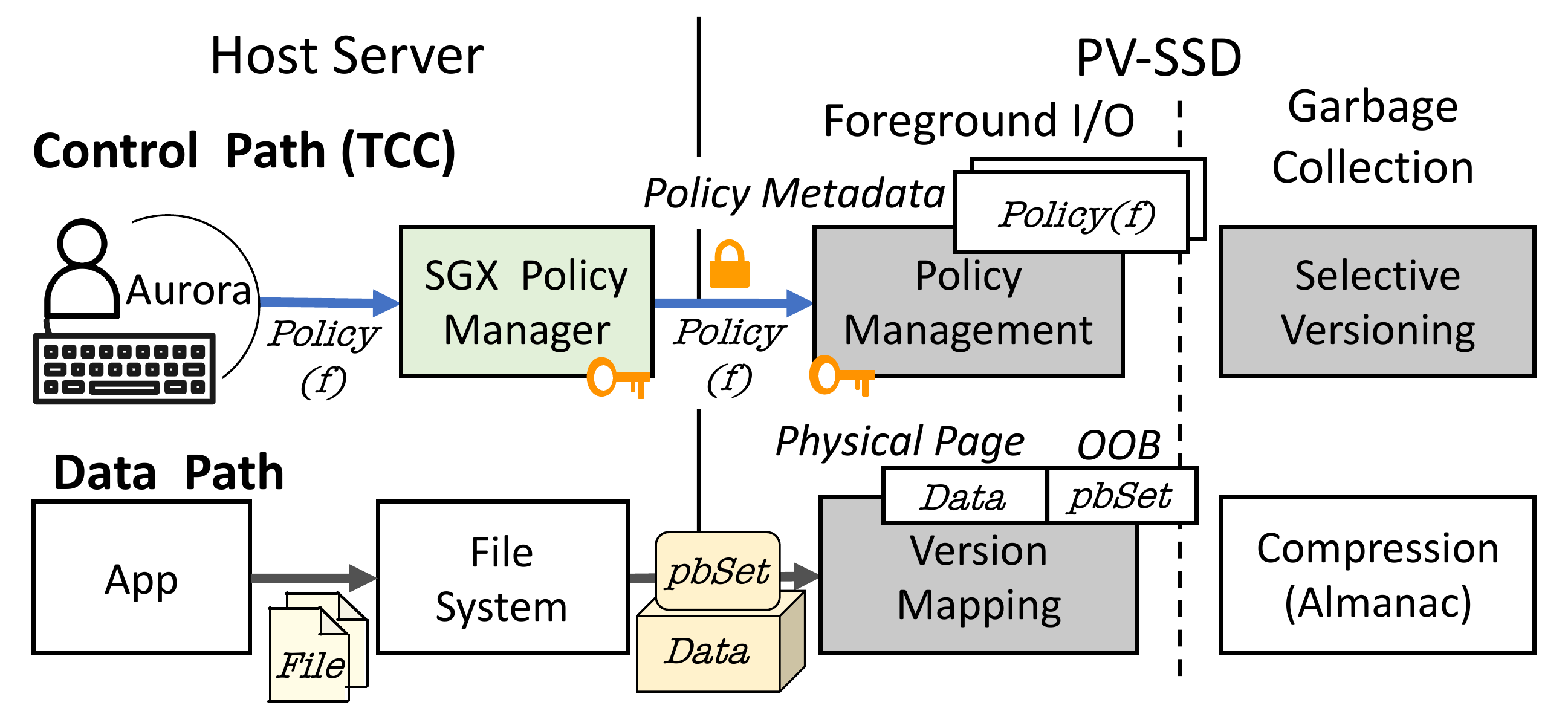} 
		\end{tabular}
		\vspace{-0.2in}
		\caption{\small An Overview of {\sgxssd}.
			}	\label{fig:overview}
		\vspace{-0.35in}
	\end{center}
\end{figure}

The user performs I/O via data path (Refer to Figure~\ref{fig:overview}). 
{\sgxssd} is designed to utilize existing native file system for allowing various applications to perform I/O transparently. 
At the same time, because \texttt{PV-SSD} does not know file semantic, so {\sgxssd} designs piggyback transfer in a way that makes \texttt{PV-SSD} aware of file-level policy. 
When writing a file, the file system sends a block I/O request to the block layer of the OS.
When the file system requests block I/O, file system piggybacks $pbSet_{(f_{j})}$ (Refer to Table~\ref{tab:notation}) into the data block.
The file path in the $pbSet_{(f_{j})}$ serves as an index for policy reference, and the file offset in the $pbSet_{(f_{j})}$ is referenced during recovery. 
This part is described in detail in Section~\ref{sec:pvssd}. 
By using piggyback transfer, the versioning information of the file for the block is transferred to \texttt{PV-SSD}.

\vspace{-0.15in}
\vspace{-0.05in}
\subsection{Policy-based File Versioning SSD}
\label{sec:pvssd}
\vspace{-0.05in}

\boldtitle{Policy Management}
Policy Management (\texttt{PM}) consists of three modules: (i) policy verifier, (ii) request handler, and (iii) policy manager. 
When \texttt{PV-SSD} receives \texttt{SPM} requests, (i) it verifies the message by comparing the MAC value using $K_{dev}$ to determine whether the request came from trusted \texttt{SPM} or not. 
Once the verification succeeds, (ii) it parses the message and extracts $P_{(f_{j})}$. 
(iii) It creates, changes or deletes the policy metadata according to the policy command type in $P_{(f_{j})}$.
The policy metadata is a hash table that is used to search for a policy entry by a file path value.
A policy entry stores the file path and CP.
When updating policy metadata, \texttt{CP} and file path in $P_{(f_{j})}$ is inserted into the policy entry.
The policy metadata is referenced by Selective Versioning (\texttt{SV}) module during Garbage Collection~\cite{SSD, alleviatingGC}.

\boldtitle{Version Mapping}
As explained in Section~\ref{sec:os}, a data block piggybacked by the file system and $pbSet_{(f_{j})}$ are transferred to Version Mapping (\texttt{VM}).
{\texttt{VM}} has three roles: (i) parsing the piggyback set ($pbSet_{(f_{j})}$), (ii) recording out-of-band (OOB) region~\cite{almanac}, and (iii) recording Version Validity Bitmap (\texttt{VVB}). 
\texttt{VVB} is a bitmap implemented in \texttt{PV-SSD} DRAM that indicates whether each physical page should be reclaimed or preserved.
A page should be preserved if the page contains \texttt{OV}.
When receiving the data written by the user, the \texttt{VM} parses the file path of $f_{j}$ and the offset of the data block from $pbSet_{(f_{j})}$.
When writing data to a NAND physical page, $pbSet_{(f_{j})}$ is recorded in the OOB region of a physical page~\cite{almanac}. 
Additionally, like Project Almanac, Logical Page Address (\texttt{LPA}), Written Time (\texttt{WT}), and Back Pointer (\texttt{BP}) -- linked to the physical page where the previous version is recorded -- are recorded in the OOB.
On the other hand, if there is no $pbSet_{(f_{j})}$ in the incoming data, nothing is written to the OOB except LPA.
In this case, there is no policy corresponding to the page, so versioning is not performed.
The \texttt{VM} sets the \texttt{VVB} bit of the page to 1 if there is a policy corresponding to the page.
\texttt{VM} resets \texttt{VVB} bit of the page to 0, if otherwise.

\boldtitle{Selective Versioning}
The \texttt{PV-SSD} selects a victim block in a greedy manner during the Garbage Collection (\texttt{GC}).
Then, in order to erase the victim block, the \texttt{PV-SSD} checks the state of each page.
If a page is in valid state, this page should be copied to another free block. 
On the other hand, if the page is not valid, the page can have two state: \texttt{Invalid Page} and Old Version Page (\texttt{OV Page}).
Here, the \texttt{OV Page} is the page containing the \texttt{OV}, so the policy-based versioning should be performed.
The main purpose of Selective Versioning (SV) is to differentiate \texttt{Invalid Page} and \texttt{OV Page} to selectively preserve only \texttt{OV}.
First, \texttt{SV} checks \texttt{VVB} to see if there is a policy corresponding to the page.
If there is no policy, the page is not required to be preserved, so it is considered as a \texttt{Invalid Page} and reclaimed during GC.
On the other hand, even if there is a corresponding policy of the page, it does not always mean that this page should be preserved.
This is because the page may have expired according to the policy.
Finally, according to \texttt{CP} recorded in the policy entry, \texttt{PV-SSD} determines whether versioning is required or not. 
If the policy of the page has expired, \texttt{PV-SSD} resets \texttt{VVB} bits corresponding to the page to 0 and reclaims the page. 
Otherwise, \texttt{PV-SSD} regards it as \texttt{OV Page}. 

\vspace{-0.15in}
\subsection{Rollback Operation}
\vspace{-0.05in}
The user can perform roll-back to a file to a certain time or a version by using the recovery tool. 
The recovery tool receives the file path of the target file ($f_j$) and the time (\texttt{T}) or the version (\texttt{V}) to restore from the user.
Then, the recovery tool uses the \texttt{ioctl()} command to load the LBA list of $f_j$ from the file system.
It sends $rec_{(f_{j})}$ (Refer to Table~\ref{tab:notation}) received from the user and the file system to the \texttt{PV-SSD} to request recovery of the file ($f_j$).
By using the LBA list of $rec_{(f_{j})}$, \texttt{PV-SSD} finds logical pages corresponding to the file contents. 
Then \texttt{PV-SSD} navigates the page chain~\cite{almanac} associated with each \texttt{LPN} to explore the \texttt{OV Page} to be recovered.
When all pages corresponding to the file contents are found, those pages are rearranged in the order of the offset recorded in the OOB area to recover the file contents.
When the restored file contents are transferred to the recovery tool, it overwrites the contents with the new file to complete recovery.
If the file system is corrupted by malware (including file metadata), the recovery tool cannot send the LBA list. 
In this case, \texttt{PV-SSD} searches all physical pages and compares the file path of $rec_{(f_{j})}$ with file path written in each page's OOB to find all corresponding LPAs to the file ($f_j$).
Then it recovers the file in the same way as described above.
In this case, it takes more time to recover, but it is possible to recover files even the file system is damaged. 


\vspace{-0.15in}
\section{Security and Performance Evaluation}
\vspace{-0.1in}

\begin{table}[!t]
	\centering	
	\begin{center}
		\resizebox{\columnwidth}{!}{
			\begin{tabular}{|p{3.8cm}|p{7cm}|}
				\hline
				\multicolumn{2}{|c|}{\textbf{Disk  Attack}}\\
				\hline
				\textbf{Attack Scenario}&\textbf{Description}\\
				\hline
				\multirow{2}{*}{(1) File Attack}&Malware attacks to tamper with users' important files.\\
				\hline
				\multirow{3}{*}{(2) Policy Deletion Attack}&The malware inserts the \texttt{DELETE} command to $P_{(f_{j})}$ and sends $P_{(f_{j})}$ to the \texttt{PV-SSD} to request policy deletion.\\
				\hline
				\multirow{2}{*}{(3) Version Attack}&Malware wears out the version of a file with continuous overwrites.\\
				\hline
				(4) File System Corruption&Malware corrupts the file system.\\
				\hline
				\hline
				\multicolumn{2}{|c|}{\textbf{Man-in-the-middle Attack }}\\
				\hline
				\textbf{Attack Scenario}&\textbf{Description}\\
				 \hline
				(5) Contents, LBA Tampering Attack&Malware tampers with the contents of the LBA of the data block that the user has requested to write.\\
				\hline
				\multirow{3}{*}{(6) Piggyback Set Deletion}&The malicious code deletes the $pbSet_{(f_{j})}$ sent during I / O so that the \texttt{PV-SSD} misunderstands that there is no policy corresponding to the data block.\\
				\hline
		\end{tabular}}
		\vspace{-0.15in}
		\caption{\small Attack Scenarios in {\sgxssd}. }
	\label{tab:attack_scenario}
	\end{center}
	\vspace{-0.4in}
\end{table}

\boldtitle{Security Analysis}
In {\sgxssd}, we analyze various attack scenarios that ring-0 level malware can attempt.
Malware attacks that compromise a host system are largely divided into two categories: (i) disk attack and (ii) man-in-the-middle attack.
(i) means that the malware directly tampers with data that is persistently stored on the disk.
On the other hand, (ii) means that the malware installs a malicious module in the kernel to falsify the message that the host server sends to the {\texttt{PV-SSD}}.
Table~\ref{tab:attack_scenario} describes various attack scenarios for each category.
(1) is a basic attack of ransomware. 
When (1) occurs, {\texttt{PV-SSD}} recovers the tampered file to the point in time before the malware invades.
In the case of (2), {\texttt{PV-SSD}} verifies $P_{(f_{j})}$ using $K_{dev}$ and denies the request.
We leave the defense of (3) at the user's discretion.
When the file priority is high, the user selectively protects the important files by not limiting the number of versions.
To recover the file under condition (4), {\texttt{PV-SSD}} searches whole physical pages, finds the pages that match with the file, reassembles them into the file, and sends it to the recovery tool.
Therefore, even in extreme situations where the file system is completely corrupted, it is possible to recover files by connecting the {\texttt{PV-SSD}} to another server.
When (5) or (6) occurs, PV-SSD cannot guarantee the integrity of newly updated data after malware intrusion.
However, since no man-in-the-middle attack has occurred for pre-existing data before malware invades, the file is recovered to the point before the malware invaded.

\begin{table}[!t]	
	
	\centering
	\resizebox{\linewidth}{!}{ 
		\begin{tabular}{|l|p{9cm}|}
			\hline
			{\bf System}& {\bf Configuration}\\
			\hline
			Host System&Intel(R) Core(TM) i7-8700 CPU @ 3.70GHz with 16 GB RAM\\
			Kernel Version&Linux 4.10.16\\
			SSD&Jasmine OpenSSD, Write/Read Throughput: 65MB/s,225MB/s \\
			\hline
	\end{tabular}}
	\vspace{-0.15in}
	\caption{\small Testbed Configurations for {\sgxssd}.}
	\label{tab:testbed_openssd}
	\vspace{-0.15in}
\end{table}

\begin{figure}[!t]
		
	\begin{center}
		\begin{tabular}{@{}c@{}c@{}c@{}c@{}}
			\includegraphics[width=0.2\textwidth]{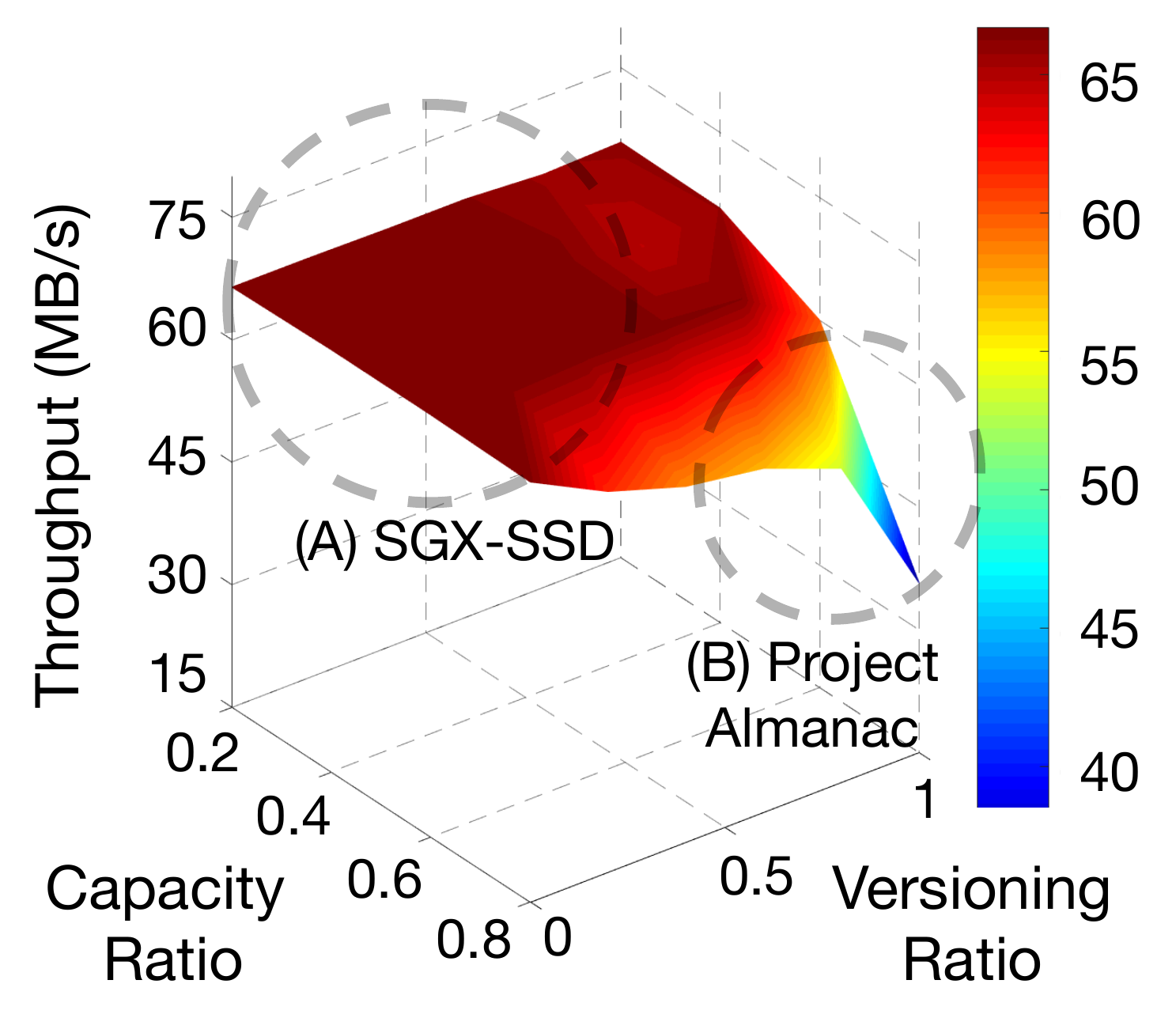} 
			&\includegraphics[width=0.2\textwidth]{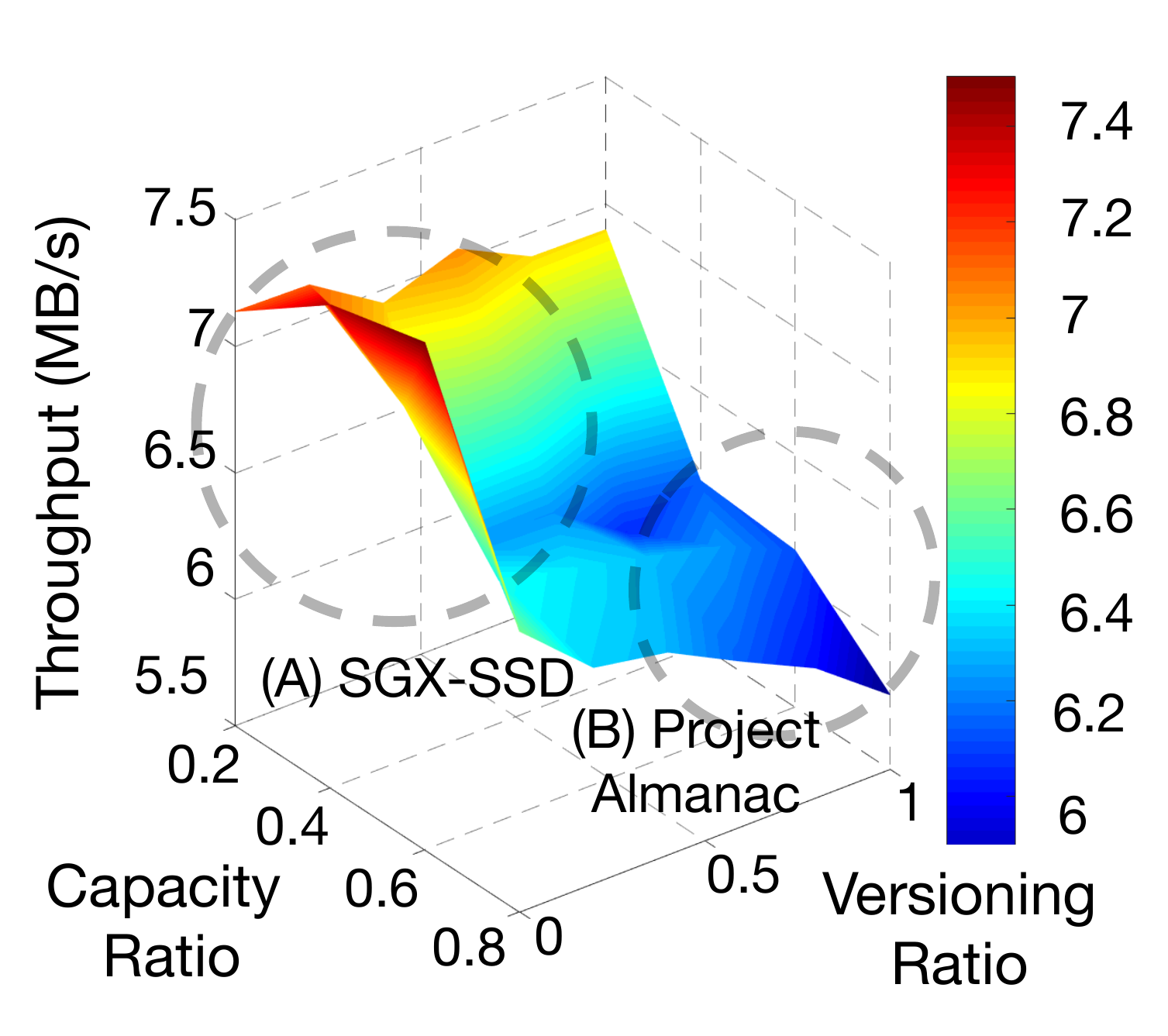}\\	
			\vspace{-0.2in}									
			{ \small (a) $Workload(B)$}
			& { \small (b) $Workload(S)$}\\
			\vspace{-0.1in}
		\end{tabular}
		\caption{\small Performance comparison of {\sgxssd} and Project Almanac.
		The capacity ratio and versioning ratio refer to the unique data ratio of the total disk capacity and the ratio of versioned data among the total data, respectively. 
			}	\label{eval:perf}
		\vspace{-0.4in}
	\end{center}
\end{figure}

\boldtitle{Performance Analysis}
For the experiment, we prototyped \texttt{PV-SSD} using Jasmine OpenSSD~\cite{jasmine} described in Table~\ref{tab:testbed_openssd}, and implemented the piggyback transfer mechanism by modifying the VFS and driver code of the kernel.
In order to induce GC in the experiment, we designated the SSD as the only partition available for 1 GB, and initialized all pages to \texttt{Invalid Page} before each experiment.
We measured the overwriting performance of synthetic Big (20MB) and small (32KB) file workloads ($Workload(B)$ and $Workload(S)$ respectively) for performance comparison according to GC overhead.
In particular, the \texttt{RT} of each data was fixed at 3 days, so that our results do not include the overhead of reclaiming the versioning data 
during the experiment period.


In Figure~\ref{eval:perf}(a)(b), we observe that {\sgxssd} maintains higher performance than Project Almanac.
Project Almanac performs versioning for all data because \texttt{RT} of every data is same.
On the other hand, {\sgxssd} selectively performs file versioning.
That is, in {\sgxssd}, only versioning of important data is performed, so write amplification is much less than Project Almanac.
The areas marked as A and B in Figure~\ref{eval:perf} show the spatial location of {\sgxssd} and Project Almanac in performance. 
In Figure~\ref{eval:perf}(b), $Workload(S)$ lowers the internal GC efficiency of SSD than $Workload(B)$, so overall throughput is low.
This is because, in the case of $Workload(S)$, \texttt{OV Page} and \texttt{Invalid Page} are mixed and distributed in the same block, thereby increasing GC overhead.
In particular, Project Almanac rapidly decreases in performance as capacity ratio and versioning ratio increase, while {\sgxssd} decreases at moderate speed.

\vspace{-0.15in}
\vspace{-0.05in}
\section{Conclusion}
\vspace{-0.1in}

In this paper, we analyze the existing state-preserving storage's vulnerability that integrity cannot be guaranteed from malware with long dwell time due to short retention time.
To solve the data integrity problem, 
we propose a policy-based file versioning storage system ({\sgxssd}). 
Through security and performance analysis, we proved that {\sgxssd} solves the existing security problems mentioned earlier and has low performance degradation due to versioning.

\vspace{-0.15in}
\section{Discussion Topics}
\vspace{-0.1in}
We would like to hear from reviewers about the following questions.

\textbf{Hardware Vulnerability}
Currently, in order to deploy {\sgxssd}, the Intel CPU with Skylake or higher that provides SGX is required, and firmware modification of the SSD is required.
We also assume SGX and the SSD are trusted.
However, the problem of vulnerability in SGX and the SSD continues to be raised~\cite{firmware_isolated,lviattack,foreshadow}.
{\sgxssd} assumes that the SSD has a timer.
However, there is no trusted timer in SSDs on the market.
This is a potential security flaw in {\sgxssd} as well as various versioning SSDs (~\cite{flashguard, almanac}).

\textbf{Challenge}
We have chosen a design where users set policies based on the importance of the files.
But, since there are so many files, is it the right design approach to pass all of these choices to the user?
Currently, we use the technology of piggyback in the OS for security stability.
However, the VFS and driver code of the OS need to be modified.
We are wondering if it will be a problem to commercialize when OS modification is needed for security.
Unlike the existing backup software~\cite{acronisbackup, extremeBinning, safeFS}, 
{\sgxssd} performs versioning on a single device, so it is impossible to cope with situations such as device failure~\cite{ssdfailure}.
Couldn't it provide better security by combining the characteristics of {\sgxssd} that protects data from high-privilege malware with existing backup software?
Currently, {\sgxssd} is designed in a stand-alone mode.
Can we expand our design and apply it to a remote storage environment where multiple users share the storage?
If applicable, what design should be added?
Should {\sgxssd} authenticate multiple users by itself?
How should each user be authorized to set up a policy and how much space should be allocated for versioning for that user?



\textbf{Originality}
Pesos~\cite{pesos} is the access-control-based object storage and provides richer policy than {\sgxssd}.
However, the security scope of Pesos is limited to remote servers, and assumes that the client system is trusted. Therefore, data cannot be protected when malware enters the client machine.
Inuksuk~\cite{inuksuk} uses Intel TXT and self-encryption disk (SED) to protect data by copying it to a protected partition.
Inuksuk, like {\sgxssd}, can selectively protect files in the disk.
However, Inuksuk has a very large overhead of backing up data to a protected partition (e.g. 23.38 seconds to back up 85.6MB of JPG files~\cite{inuksuk}), and the whole system is interrupted while data is being backed up.
Due to the long system downtime problem, Inuksuk is unable to back up data in real time, and the size of the files to be backed up and the backup cycle is limited.
In particular, in the client-server model, where service interruption is very sensitive, an Inuksuk server must be additionally deployed to prevent service interruption.
On the other hand, {\sgxssd} not only performs host service and {\texttt{PV-SSD}}'s versioning at the same time, but also has very little overhead caused by versioning. 

\vspace{-0.15in}

\subsection*{Acknowledgments}
\vspace{-0.05in}
This work was supported by Samsung Semiconductor research grant.

\vspace{-0.2in}

\small
\bibliographystyle{plain}
\bibliography{paper}

\end{document}